\documentclass[aps,prl,superscriptaddress,amsfonts,amsmath,amssymb,floatfix,reprint,longbibliography]{revtex4-2}

\usepackage{amsmath}
\usepackage{amssymb}
\usepackage{xcolor}
\usepackage[colorlinks=true, urlcolor=blue, linkcolor=blue, citecolor=teal, pdftex]{hyperref}
\usepackage{multirow}
\usepackage{natbib}
\usepackage{siunitx}
\usepackage{tikz}
\usepackage[normalem]{ulem}
\usepackage{csquotes}
\usepackage[version=4]{mhchem}
\usepackage{float}

\usepackage{braket}
\usepackage{times}

\usepackage{pgfplots}
\usepackage{pgfplotstable}
\pgfplotsset{width=\columnwidth,compat=1.9}
\usepgfplotslibrary{external}
\usepgfplotslibrary{colormaps}

\usetikzlibrary{patterns,shapes.arrows,arrows,fadings,decorations.pathmorphing}
\usetikzlibrary{shapes.geometric,patterns,positioning,matrix}
\usetikzlibrary{shadows,math,calc,3d,perspective,arrows.meta,decorations.pathmorphing,decorations.markings,decorations.pathreplacing}
\usetikzlibrary{backgrounds}

\usepackage{tabularray}

\DeclareSIUnit{\million}{\text{million}}

\definecolor{plot1}{RGB}{31, 119, 180}
\definecolor{plot2}{RGB}{255, 127, 14}
\definecolor{plot3}{RGB}{44, 160, 44}
\definecolor{plot4}{RGB}{214, 39, 40}
\definecolor{plot5}{RGB}{148, 103, 189}
\definecolor{plot6}{RGB}{140, 86, 75}
\definecolor{plot7}{RGB}{227, 119, 194}
\definecolor{plot8}{RGB}{127, 127, 127}
\definecolor{plot9}{RGB}{188, 189, 34}
\definecolor{plot10}{RGB}{23, 190, 207}

\definecolor{mathematicaplot1}{rgb}{0.368417, 0.506779, 0.709798}
\definecolor{mathematicaplot2}{rgb}{0.880722, 0.611041, 0.142051}
\definecolor{mathematicaplot3}{rgb}{0.560181, 0.691569, 0.194885}
\definecolor{mathematicaplot4}{rgb}{0.922526, 0.385626, 0.209179}
\definecolor{mathematicaplot5}{rgb}{0.528488, 0.470624, 0.701351}
\definecolor{mathematicaplot6}{rgb}{0.772079, 0.431554, 0.102387}
\definecolor{mathematicaplot7}{rgb}{0.363898, 0.618501, 0.782349}
\definecolor{mathematicaplot8}{rgb}{1, 0.75, 0}
\definecolor{mathematicaplot9}{rgb}{0.647624, 0.37816, 0.614037}
\definecolor{mathematicaplot10}{rgb}{0.571589, 0.586483, 0.}
\definecolor{mathematicaplot11}{rgb}{0.915, 0.3325, 0.2125}
\definecolor{mathematicaplot12}{rgb}{0.400822, 0.522007, 0.85}
\definecolor{mathematicaplot13}{rgb}{0.972829, 0.621644, 0.073362}
\definecolor{mathematicaplot14}{rgb}{0.736783, 0.358, 0.503027}
\definecolor{mathematicaplot15}{rgb}{0.280264, 0.715, 0.429209}

\definecolor{googleB}{HTML}{4285F4}
\definecolor{googleG}{HTML}{34A853}
\definecolor{googleY}{HTML}{FBBC05}
\definecolor{googleR}{HTML}{EA4335}
\definecolor{googleBG}{HTML}{3B96A4}

\pgfplotsset{%
    colormap={bluewhite}{
        rgb255=(8,48,107)
        rgb255=(32,112,180)
        rgb255=(106,173,213)
        rgb255=(197,218,238)
        rgb255=(247,251,255)
    },
}

\begin{document}

\title{An extended non-magnetic phase in the spin-1/2 Heisenberg antiferromagnet\\
from the ruby to the maple-leaf lattice}

\newcommand{\FUB}{Freie Universität Berlin, Dahlem Center for Complex Quantum Systems and Institut für Theoretische Physik,  Arnimallee 14, 14195 Berlin, Germany}

\newcommand{\HZB}{Helmholtz-Zentrum Berlin für Materialien und Energie, Hahn-Meitner-Platz 1, 14109 Berlin, Germany}

\newcommand{\UOC}{University of Cologne, Institute for Theoretical Physics, Zülpicher Straße 77, 50937 Köln, Germany}

\newcommand{\IIT}{Indian Institute of Technology Madras, Department of Physics and Quantum Centre of Excellence for Diamond and Emergent Materials (QuCenDiEM), Chennai 600036, India}

\author{Philipp Schmoll}
\thanks{Both authors contributed equally.}
\affiliation{\FUB}

\author{Jan Naumann}
\thanks{Both authors contributed equally.}
\affiliation{\FUB}

\author{Erik L. Weerda}
\affiliation{\UOC}

\author{Jens Eisert}
\affiliation{\FUB}
\affiliation{\HZB}

\author{Yasir Iqbal}
\affiliation{\IIT}


\begin{abstract}
The spin-$1/2$ Heisenberg antiferromagnet on the two-dimensional ruby and maple-leaf lattices provides a stringent test case for frustrated quantum magnetism, where semiclassical magnetic order competes closely with quantum-paramagnetic states.
We study the generalized maple-leaf model along the axis interpolating between the isotropic ruby lattice and the isotropic maple-leaf lattice, using two independent variational infinite projected entangled-pair state ans\"atze: one defined on the mapped square lattice and one directly on the native triangular lattice.
Our lowest-energy variational states show no magnetic order throughout the region connecting the two isotropic limits, and the local bond correlations recover the $120^\circ$ lattice rotation symmetry without imposing it explicitly.
Magnetic-field simulations at the two endpoints further indicate a finite zero-magnetization plateau in both cases, with a gradual onset of magnetization on the ruby lattice and a sharper onset on the maple-leaf lattice.
These results establish a broad gapped non-magnetic regime in a minimal nearest-neighbor Heisenberg model on lattices of direct relevance to both frustrated quantum magnets and programmable ruby-lattice simulators.
\end{abstract}

\maketitle

\emph{Quantum spin liquids} (QSLs) are quantum-disordered magnets in which the absence of conventional order is accompanied by more stringent forms of many-body organization, such as long-range entanglement and fractionalized quasiparticle excitations~\cite{Balents2010,Savary2017}. This distinction is important: a vanishing magnetic order parameter by itself does not establish a QSL. A non-magnetic ground state may instead be a valence-bond crystal, a short-range resonating-valence-bond state with unresolved lattice symmetry breaking, or even a fully featureless paramagnet when the lattice filling and crystalline quantum numbers do not impose a Lieb-Schultz-Mattis-type obstruction~\cite{Po-2017}. The numerical problem is therefore twofold: one must first decide whether magnetic order survives in the thermodynamic limit, and then determine whether the resulting paramagnet is topologically ordered or conventional. A useful benchmark is the kagome Heisenberg antiferromagnet, where the ground state is widely accepted to be non-magnetic, while the distinction between a gapped $\mathbb Z_2$ spin liquid~\cite{Yan2011,Depenbrock2012,Laeuchli2019} and a gapless $U(1)$ Dirac spin liquid~\cite{Iqbal-2013,Iqbal-2014,Liao2017,He2017} has remained numerically delicate.

\tikzexternaldisable
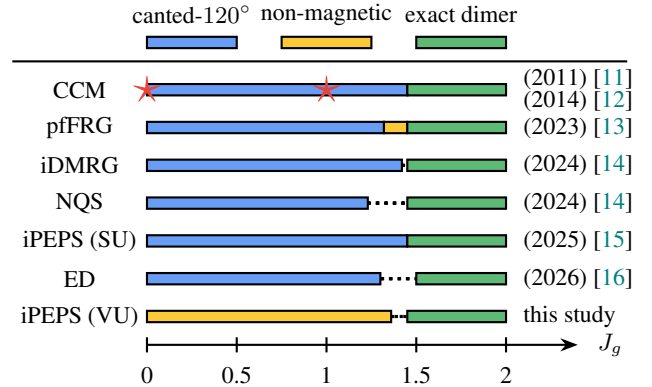
\begin{figure}[b]
  \centering
  \begin{tikzpicture}
    \def\F{2.375}
    \def\boxSize{0.075}
    \def\yShift{0.5}

    \draw[thick, fill = googleB!80, shift = {(+0.25 * \F, +3.25 * \yShift)}] (-0.25 * \F, -\boxSize) rectangle (+0.25 * \F, +\boxSize);
    \node[align = center] at (+0.25 * \F, +4.00 * \yShift) {canted-$120^\circ$};

    \draw[thick, fill = googleY!80, shift = {(+1.0 * \F, +3.25 * \yShift)}] (-0.25 * \F, -\boxSize) rectangle (+0.25 * \F, +\boxSize);
    \node[align = center] at (+1.00 * \F, +4.00 * \yShift) {non-magnetic};

    \draw[thick, fill = googleG!80, shift = {(1.75 * \F, +3.25 * \yShift)}] (-0.25 * \F, -\boxSize) rectangle (+0.25 * \F, +\boxSize);
    \node[align = center] at (+1.75 * \F, +4.00 * \yShift) {exact dimer};

    \draw[thick, shift = {(0, +2.75 * \yShift)}] (-0.75 * \F, 0) to (+2.675 * \F, 0);

    \begin{scope}[shift = {(0, +2*\yShift)}]
      \node at (-0.375 * \F, 0) {CCM};
      \node[anchor=west, align=center] at (+2.05 * \F, 0) {(2011)~\cite{Farnell2011}\\[-0.25em](2014)~\cite{Farnell2014}};
      \draw[thick, fill = googleB!80] (0.00 * \F, -\boxSize) rectangle (1.45 * \F, +\boxSize);
      \draw[thick, fill = googleG!80] (1.45 * \F, -\boxSize) rectangle (2.00 * \F, +\boxSize);
      \node[googleR] at (0.0 * \F, 0) {\huge$\star$};
      \node[googleR] at (1.0 * \F, 0) {\huge$\star$};
    \end{scope}

    \begin{scope}[shift = {(0, +1*\yShift)}]
      \node at (-0.375 * \F, 0) {pfFRG};
      \node[anchor=west] at (+2.05 * \F, 0) {(2023)~\cite{Gresista2023}};
      \draw[thick, fill = googleB!80] (0.00 * \F, -\boxSize) rectangle (1.32 * \F, +\boxSize);
      \draw[thick, fill = googleY!80] (1.32 * \F, -\boxSize) rectangle (1.45 * \F, +\boxSize);
      \draw[thick, fill = googleG!80] (1.45 * \F, -\boxSize) rectangle (2.00 * \F, +\boxSize);
    \end{scope}

    \begin{scope}[shift = {(0, 0)}]
      \node at (-0.375 * \F, 0) {iDMRG};
      \node[anchor=west] at (+2.05 * \F, 0) {(2024)~\cite{Beck2024}};
      \draw[thick, fill = googleB!80] (0.00 * \F, -\boxSize) rectangle (1.42 * \F, +\boxSize);
      \draw[very thick, dotted] (1.42 * \F, 0) rectangle (1.45 * \F, 0);
      \draw[thick, fill = googleG!80] (1.45 * \F, -\boxSize) rectangle (2.00 * \F, +\boxSize);
    \end{scope}

    \begin{scope}[shift = {(0, -1*\yShift)}]
      \node at (-0.375 * \F, 0) {NQS};
      \node[anchor=west] at (+2.05 * \F, 0) {(2024)~\cite{Beck2024}};
      \draw[thick, fill = googleB!80] (0.00 * \F, -\boxSize) rectangle (1.23 * \F, +\boxSize);
      \draw[very thick, dotted] (1.23 * \F, 0) rectangle (1.45 * \F, 0);
      \draw[thick, fill = googleG!80] (1.45 * \F, -\boxSize) rectangle (2.00 * \F, +\boxSize);
    \end{scope}

    \begin{scope}[shift = {(0, -2*\yShift)}]
      \node at (-0.375 * \F, 0) {iPEPS (SU)};
      \node[anchor=west] at (+2.05 * \F, 0) {(2025)~\cite{Nyckees2025}};
      \draw[thick, fill = googleB!80] (0.00 * \F, -\boxSize) rectangle (1.45 * \F, +\boxSize);
      \draw[thick, fill = googleG!80] (1.45 * \F, -\boxSize) rectangle (2.00 * \F, +\boxSize);
    \end{scope}

    \begin{scope}[shift = {(0, -3*\yShift)}]
      \node at (-0.375 * \F, 0) {ED};
      \node[anchor=west] at (+2.05 * \F, 0) {(2026)~\cite{Ebert2026}};
      \draw[thick, fill = googleB!80] (0.00 * \F, -\boxSize) rectangle (1.30 * \F, +\boxSize);
      \draw[very thick, dotted] (1.30 * \F, 0) rectangle (1.50 * \F, 0);
      \draw[thick, fill = googleG!80] (1.50 * \F, -\boxSize) rectangle (2.00 * \F, +\boxSize);
    \end{scope}

    \begin{scope}[shift = {(0, -4*\yShift)}]
      \node at (-0.375 * \F, 0) {iPEPS (VU)};
      \node[anchor=west] at (+2.05 * \F, 0) {\text{this study}};
      \draw[thick, fill = googleY!80] (0.00 * \F, -\boxSize) rectangle (1.36 * \F, +\boxSize);
      \draw[very thick, dotted] (1.36 * \F, 0) rectangle (1.45 * \F, 0);
      \draw[thick, fill = googleG!80] (1.45 * \F, -\boxSize) rectangle (2.00 * \F, +\boxSize);
    \end{scope}

    \begin{scope}[shift = {(0, -4.75*\yShift)}]
      \node[right = 0.15] at (+2.4 * \F, 0) {$J_g$};
      \draw[thick, black, -{Stealth[scale = 1.0]}] (0.00 * \F, 0) -- (+2.4 * \F, 0);
      \draw[thick, shift = ({0.0 * \F, 0})] (0, -0.1) to (0, +0.1) node at (0, -0.4) {0};
      \draw[thick, shift = ({0.5 * \F, 0})] (0, -0.1) to (0, +0.1) node at (0, -0.4) {0.5};
      \draw[thick, shift = ({1.0 * \F, 0})] (0, -0.1) to (0, +0.1) node at (0, -0.4) {1};
      \draw[thick, shift = ({1.5 * \F, 0})] (0, -0.1) to (0, +0.1) node at (0, -0.4) {1.5};
      \draw[thick, shift = ({2.0 * \F, 0})] (0, -0.1) to (0, +0.1) node at (0, -0.4) {2};
    \end{scope}
  \end{tikzpicture}
  \caption{One-dimensional phase diagram of the purely antiferromagnetic, generalized maple-leaf Heisenberg model with \mbox{$J_{r} = J_{b} = 1$} and \mbox{$J_{g} \in \lbrack 0, 2 \rbrack$}, as reported by several numerical methods in the literature. The dual variational PEPS results reported here identify a broad non-magnetic regime connecting the ruby and MLL limits. Red stars denote the possibility of vanishing magnetic order depending on the extrapolation scheme used~\cite{Farnell2014}, dashed lines indicate inconclusive regions~\cite{Beck2024}. Figure adapted from an image originally published in Ref.~\cite{Nyckees2025}.}
  \label{fig1}
\end{figure}
\tikzexternalenable

\begin{figure}[ht]
    \centering
    \includegraphics[scale = 0.9]{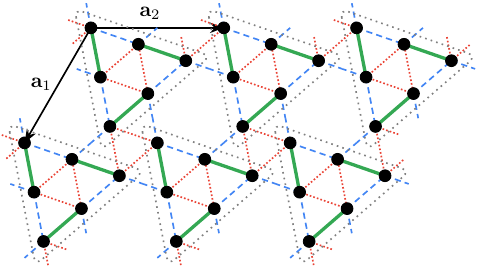}
    \caption{Illustration of the maple-leaf lattice with the symmetry inequivalent bonds in red, green and blue. In the absence of green bonds, one obtains the ruby lattice. Both lattices have a six-site geometric unit cell basis (shown by the dotted gray triangles) with an underlying triangular Bravais lattice spanned by ${\bf a}_{1}$ and ${\bf a}_{2}$.}
    \label{fig:mapleLeafLattice_0}
\end{figure}

Recently, the less-explored two-dimensional \emph{ruby lattice}~\cite{Richter2004} and \emph{maple-leaf lattice} (MLL)~\cite{Betts-1995} [see Fig.~\ref{fig:mapleLeafLattice_0}] have attracted growing theoretical, experimental, and synthetic-platform interest.
The MLL is a regular $1/7$ site-depleted triangular lattice with coordination number five, intermediate between the kagome and triangular lattices, and contains three symmetry-inequivalent nearest-neighbor bonds.
This structure provides a simple route to strong geometric frustration and enhanced quantum fluctuations for antiferromagnetic interactions.
By tuning one of the three couplings to zero, the MLL can be continuously connected to the ruby lattice, which may be viewed as a bond-depleted MLL with coordination number four and one frustrating bond removed.
The resulting interpolation therefore provides a minimal setting in which to study how connectivity and frustration control the stability of non-magnetic quantum states.

This connection has become especially timely in light of recent progress in programmable quantum simulation.
Ruby-lattice Rydberg arrays have provided a route to probing toric-code-type $\mathbb Z_2$ topological order under precisely controlled conditions and to direct measurements of nonlocal string diagnostics~\cite{
Semeghini-2021,Veressen-2021,Giudici-2022,Veressen-2022,Glaetzle-2014,Samajdar-2023,Maity-2024-2}.
At the same time, recent Floquet-engineering proposals suggest that tunable Heisenberg and extended-Heisenberg models may be realizable on a broad class of programmable Archimedean lattices~\cite{Tian2025}.
The ruby point studied here should therefore be viewed not only as a limiting case of the generalized MLL Heisenberg model, but also as a clean $SU(2)$-symmetric counterpart to the ruby-lattice spin-liquid physics now being explored in synthetic platforms.
On the materials side, the MLL underlies several copper-based compounds,
including spangolite~\cite{Hawthorne1993,Schmoll2024}, sabelliite~\cite{Olmi1995}, mojaveite~\cite{Mills2014}, fuetterite~\cite{Kampf2013}, and bluebellite~\cite{Mills2014,Makuta-2021,Ghosh2023}, as well as more semiclassical antiferromagnets such as
MgMn$_3$O$_7$$\cdot$3H$_2$O~\cite{Haraguchi2018} and Na$_2$Mn$_3$O$_7$~\cite{Venkatesh2020,Saha-2023}.

Motivated by this state of affairs, we explore the ground state phase diagram of the Heisenberg Hamiltonian along a single parameter axis, interpolating between the isotropic ruby and MLL points. We refer to this model as the \textit{generalized} MLL model.
While a central debate in the literature concerns whether the ground state of the isotropic Heisenberg model is a quantum paramagnet or features long-range magnetic order~\cite{Schmalfuss2002,Farnell2014,Farnell2011,Richter2004}, the phase diagram of the generalized MLL model has also been subject to considerable research effort, some of which we summarize in Fig.~\ref{fig1}.
However, studies based on \emph{exact diagonalization} (ED) and \emph{coupled cluster methods} (CCM)~\cite{Schmalfuss2002,Farnell2011,Farnell2014}, infinite \emph{density-matrix renormalization group} (DMRG) and \emph{neural quantum states} (NQS)~\cite{Ghosh2022,Beck2024,Ghosh2024}, as well as \emph{pseudo-fermion functional renormalization group} (pf-FRG)~\cite{Gresista2023} approaches have reached varying conclusions.
Several calculations favour a canted-$120^\circ$ magnetically ordered state over much of this axis~\cite{Schmalfuss2002,Farnell2011,Beck2024,Ghosh2024}, but the evidence is not uniform.
In CCM, the extrapolated order parameter depends sensitively on the empirical extrapolation scheme and can be driven to zero~\cite{Farnell2014}.
pf-FRG finds a candidate quantum-disordered intermediate regime~\cite{Gresista2023}.
More recently, NLCE results for the isotropic MLL point converge to zero temperature and point to a short-range correlated quantum paramagnet built from resonating hexagonal motifs~\cite{Schaefer2026}.
As the interpolation parameter increases beyond the MLL point in the generalized model, CCM calculations predict a direct transition from the canted-$120^\circ$ order to a dimer product state~\cite{Schmalfuss2002,Farnell2011}, supported by a recent infinite PEPS~\cite{Nyckees2025} and ED study~\cite{Ebert2026}.
Conversely, pf-FRG calculations~\cite{Gresista2023} suggest the existence of an intermediate quantum paramagnetic phase, which may align with a narrow inconclusive region reported in the iDMRG and NQS investigations~\cite{Beck2024}.

Indeed, treating frustrated quantum systems numerically poses significant technical and computational challenges.
First, the exponential many-body Hilbert space limits exact calculations to small system sizes, giving rise to large finite size effects that complicate the extraction of information about the thermodynamic limit.
Second, frustration typically leads to strong energetic competition between different phases and complex, possibly long-range entanglement patterns in the system.
\emph{Tensor networks} (TNs), building on area laws for entanglement entropies~\cite{AreaReview}, provide a powerful toolbox for the theoretical and numerical study of quantum many-body systems~\cite{Orus2019,Cirac2021}, with method development still ongoing. 
The two-dimensional version of TNs~\cite{Verstraete2004}, the (infinite) \emph{projected entangled-pair state} (PEPS), has technically matured substantially in recent years due to the development of variational optimization techniques for energy minimization~\cite{PhysRevB.94.035133,PhysRevB.94.155123,PhysRevX.9.031041}.
This makes them highly competitive with other numerical techniques, such as variational quantum Monte Carlo, density-matrix renormalization group or coupled cluster methods.

In this Letter, we address the delicate question of whether magnetic order survives in the ground state of the generalized MLL model.
Our results indicate the absence of magnetic order at the ruby and MLL limits, and throughout the region interpolating between them.
We also find no evidence for translation or point-group symmetry breaking within our variational ansatz, and finite-field simulations at the two isotropic endpoints show zero-magnetization plateaus consistent with finite spin gaps.
We therefore identify an extended symmetric quantum-paramagnetic regime. We deliberately use this terminology in the main text: the present diagnostics establish a strong case against conventional magnetic order, but they do not by themselves distinguish a topologically ordered QSL from a featureless or very weakly symmetry-breaking short-range paramagnet.
Our results highlight the substantial challenges that complex frustrated magnets can present to numerical methods, where finite size, finite entanglement, or additional approximations may bias the calculation toward magnetically ordered states.\\

\paragraph{Model and methods.}
We study the spin-$1/2$ Heisenberg model on the  MLL with the Hamiltonian
\begin{align}
    \mathcal H = J_r \mkern-4mu \sum_{\langle i,j \rangle_r} \mathbf S_{i} \cdot \mathbf S_{j} + J_g \mkern-4mu \sum_{\langle i,j \rangle_g} \mathbf S_{i} \cdot \mathbf S_{j} + J_b \mkern-4mu \sum_{\langle i,j \rangle_b} \mathbf S_{i} \cdot \mathbf S_{j},
    \label{eq:modelDefinitionMLM}
\end{align}
where $\langle i,j \rangle$ denotes the three types of symmetry inequivalent nearest-neighbor bonds, visualized as red, green and blue lines in Fig.~\ref{fig:mapleLeafLattice_0}.
For simulations in the presence of an external magnetic field, the Hamiltonian in Eq.~\eqref{eq:modelDefinitionMLM} is extended by an additional Zeeman term $-h_z \sum_{i} S_{i}^z$.
We focus on the fully antiferromagnetic model with two of the couplings fixed as $J_r = J_b = 1$ and varying the interaction strength $J_g \in [0,1.5]$ on green bonds.
Thus our study includes both the isotropic ruby lattice ($J_g = 0$) and MLL ($J_g = 1$) limits, and tunes into the exact dimer phase for large $J_g$, for which spins on the green bonds pair into singlets~\cite{Misguich-1999,Ghosh2022}.

For the simulation of the Hamiltonian in Eq.~\eqref{eq:modelDefinitionMLM}, we employ variational infinite PEPS in two complementary geometries.
The six spins in the elementary unit cell of the MLL are first coarse-grained into an effective site on the underlying triangular Bravais lattice, spanned by $\mathbf{a}_1$ and $\mathbf{a}_2$ as indicated in Fig.~\ref{fig:mapleLeafLattice_0}.
We then treat this coarse-grained lattice either as a square-lattice PEPS with one diagonal interaction $\mathbf{a}_1+\mathbf{a}_2$, or directly as a native triangular-lattice PEPS.
The latter representation treats the three lattice directions on equal footing and is better adapted to the entanglement geometry of the generalized MLL~\cite{Naumann2026}.

In both cases we use the \emph{spiral PEPS ansatz}~\cite{Hasik2023} in which a translation-invariant PEPS tensor is combined with a spatially dependent unitary rotation parameterized by a wave vector
$\mathbf{k}\in\mathbb{R}^2$.
The ordering wave vector is therefore optimized variationally together with the tensor, rather than imposed through a large
magnetic unit cell. The energy minimization is performed using automatic differentiation~\cite{Naumann2023}, giving fully variational thermodynamic-limit wave functions at each finite bond dimension $\chi_B$.
This is important for the present work: the non-magnetic states discussed below are obtained directly as optimized finite-$\chi_B$ variational wave functions, not only through an extrapolation of the magnetic order parameter.
To achieve sufficient accuracy, the environment bond dimension chosen for the optimization is set up to $\chi_E = 6\chi_B^2$.
For further details on the numerical setup, we refer to the End Matter and to Refs.~\cite{Naumann2026,Hasik2023,Silvi2019,Schmoll2020,Chubukov-1984,Iqbal-2016,Nishino1996,Nishino1997,Orus2009}.\\

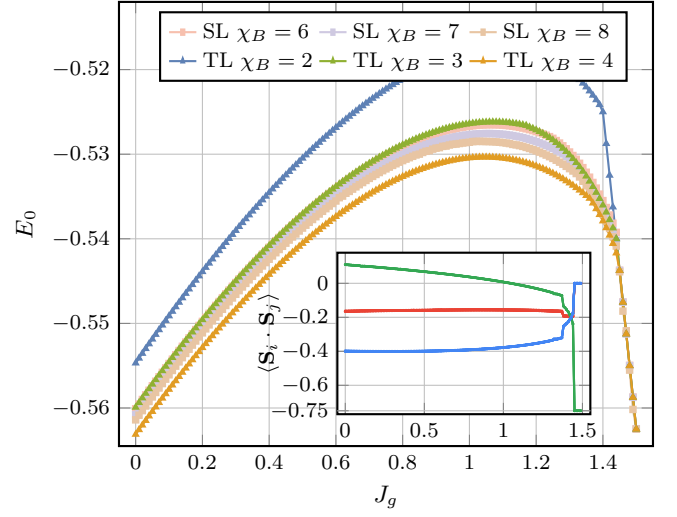
\begin{figure}[t]
    \centering
    \tikzsetnextfilename{phaseDiagram_spiralPEPS_GSE_TL}
    \begin{tikzpicture}
    
    \begin{axis}[
        xlabel=$J_g$,
        ylabel=$E_0$,
        width=\columnwidth,
        legend style={
            font=\footnotesize,
            at={(0.075, 0.82)},
            anchor=south west
        },
        legend image post style={xscale=0.6},
        legend columns=3,
        legend cell align={left},
        label style={font=\small},
        tick label style={font=\footnotesize},
        every axis/.append style={thick},
        /tikz/mark size=1pt,
        xmajorgrids=true,
        ymajorgrids=true,
        xmin=-0.05, xmax=1.55,
        ymin=-0.5645, ymax=-0.512
    ]

    \addplot[mathematicaplot4!40!white, mark=square*] table [x=Jg,y=D6] {data/phaseDiagram_spiralPEPS_GSE.txt};
    \addlegendentry{SL $\chi_B = 6$}

    \addplot[mathematicaplot5!40!white, mark=square*] table [x=Jg,y=D7] {data/phaseDiagram_spiralPEPS_GSE.txt};
    \addlegendentry{SL $\chi_B = 7$}

    \addplot[mathematicaplot6!40!white, mark=square*] table [x=Jg,y=D8] {data/phaseDiagram_spiralPEPS_GSE.txt};
    \addlegendentry{SL $\chi_B = 8$}

    \addplot[mathematicaplot1, mark=triangle*] table [x=Jg,y=D2] {data/phaseDiagram_spiralPEPS_GSE_TL.txt};
    \addlegendentry{TL $\chi_B = 2$}

    \addplot[mathematicaplot3, mark=triangle*] table [x=Jg,y=D3] {data/phaseDiagram_spiralPEPS_GSE_TL.txt};
    \addlegendentry{TL $\chi_B = 3$}

    \addplot[mathematicaplot2, mark=triangle*] table [x=Jg,y=D4] {data/phaseDiagram_spiralPEPS_GSE_TL.txt};
    \addlegendentry{TL $\chi_B = 4$}
    
    \end{axis}
  
    \begin{axis}[
        axis background/.style={fill=white},
        ylabel=$\langle \mathbf{S}_i \cdot\mathbf{S}_j \rangle$,
        width=0.57\columnwidth,
        height=0.43\columnwidth,
        legend style={
            font=\footnotesize, 
            at={(0.07, 0.01)},
            anchor=south west
        },
        legend image post style={xscale=0.4},
        legend cell align={left},
        label style={font=\footnotesize},
        tick label style={font=\footnotesize},
        every axis/.append style={thick},
        clip mode=individual,
        /tikz/mark size=0.1pt,
        xmajorgrids=true,
        ymajorgrids=true,
        xshift=82.2,
        yshift=12,
        xlabel shift=-9,
        ylabel shift=-10,
        xmin=-0.05, xmax=1.55,
        ymin=-0.77, ymax=0.18,
        extra y ticks={-0.75},
    ]

    \addplot[googleR, mark=*, forget plot] table [x=Jg,y=S2S4] {data/phaseDiagram_spiralPEPS_indSpinCorr_chiB_4_TL.txt};
    \addplot[googleR, mark=*, forget plot] table [x=Jg,y=S2S5] {data/phaseDiagram_spiralPEPS_indSpinCorr_chiB_4_TL.txt};
    \addplot[googleR, mark=*, forget plot] table [x=Jg,y=S4S5] {data/phaseDiagram_spiralPEPS_indSpinCorr_chiB_4_TL.txt};
    \addplot[googleR, mark=*, forget plot] table [x=Jg,y=S3S6] {data/phaseDiagram_spiralPEPS_indSpinCorr_chiB_4_TL.txt};
    \addplot[googleR, mark=*, forget plot] table [x=Jg,y=S6S1] {data/phaseDiagram_spiralPEPS_indSpinCorr_chiB_4_TL.txt};
    \addplot[googleR, mark=*] table [x=Jg,y=S3S1] {data/phaseDiagram_spiralPEPS_indSpinCorr_chiB_4_TL.txt};

    \addplot[googleG, mark=*, forget plot] table [x=Jg,y=S1S2] {data/phaseDiagram_spiralPEPS_indSpinCorr_chiB_4_TL.txt};
    \addplot[googleG, mark=*, forget plot] table [x=Jg,y=S3S4] {data/phaseDiagram_spiralPEPS_indSpinCorr_chiB_4_TL.txt};
    \addplot[googleG, mark=*] table [x=Jg,y=S5S6] {data/phaseDiagram_spiralPEPS_indSpinCorr_chiB_4_TL.txt};

    \addplot[googleB, mark=*, forget plot] table [x=Jg,y=S1S5] {data/phaseDiagram_spiralPEPS_indSpinCorr_chiB_4_TL.txt};
    \addplot[googleB, mark=*, forget plot] table [x=Jg,y=S2S3] {data/phaseDiagram_spiralPEPS_indSpinCorr_chiB_4_TL.txt};
    \addplot[googleB, mark=*, forget plot] table [x=Jg,y=S4S6] {data/phaseDiagram_spiralPEPS_indSpinCorr_chiB_4_TL.txt};
    \addplot[googleB, mark=*, forget plot] table [x=Jg,y=S3S5] {data/phaseDiagram_spiralPEPS_indSpinCorr_chiB_4_TL.txt};
    \addplot[googleB, mark=*, forget plot] table [x=Jg,y=S6S2] {data/phaseDiagram_spiralPEPS_indSpinCorr_chiB_4_TL.txt};
    \addplot[googleB, mark=*] table [x=Jg,y=S4S1] {data/phaseDiagram_spiralPEPS_indSpinCorr_chiB_4_TL.txt};
    
    \end{axis}

\end{tikzpicture}
    \caption{Ground state energy per spin as a function of the coupling $J_g$ at various bond dimensions $\chi_{B}$ for PEPS defined on the square lattice (SL) and on the native triangular lattice (TL). The linear slope for \mbox{$J_g > 1.45$} is given by $E_0 = -3J_g/8$, i.e., the energy of the exact dimer state. The inset shows the individual bond energies \mbox{$\langle \mathbf{S}_i\cdot\mathbf{S}_j \rangle$} for the fifteen different nearest-neighbor bonds in the elementary unit cell (TL, $\chi_B = 4$), colored according to the three inequivalent types of bond in the lattice (cf.\ Fig.~\ref{fig:mapleLeafLattice_0}). We observe a numerically accurate $120^\circ$ rotational symmetry, without this symmetry being explicitly imposed on the PEPS tensor.}
    \label{fig:phaseDiagram_spiralPEPS_GSE_TL}
\end{figure}

\paragraph{Results.}
With the chosen setup of Heisenberg interactions, we map out a one-dimensional slice of the antiferromagnetic region of the phase diagram.
In Fig.~\ref{fig:phaseDiagram_spiralPEPS_GSE_TL}, we present the variational ground state energies per spin $E_{0}$ as a function of the coupling parameter $J_g$ for both the infinite square lattice (SL) and triangular lattice (TL) PEPS ansätze.
Increasing the bulk bond dimension $\chi_B$ leads to a consistent decrease in energy in both cases, as more entanglement is captured in the system.
The point of maximal frustration is located in the vicinity of the isotropic MLL antiferromagnet, where $E_0$ reaches its highest value and the system is least able to minimize each local Hamiltonian term simultaneously.
The progressive decrease in $E_{0}$, and hence in frustration, as $J_{g}$ is lowered, reflects the fact that the system loses a frustrating bond as it approaches the ruby lattice limit, despite its lower coordination number $z=4$ compared to $z=5$ for the MLL.
This may explain the comparatively higher sensitivity of $E_{0}$ to $\chi_{B}$ with increasing $J_{g}$.
The known exact dimer phase is found for $J_{g} > 1.45$~\cite{Farnell2011,Gresista2023,Beck2024}, where all simulations collapse to the analytical energy of $E_{0} = -3J_{g}/8$, with $\langle \mathbf S_{i}\cdot\mathbf S_{j} \rangle = -3/4$ on the green bonds and zero on the red and blue bonds [see inset of Fig.~\ref{fig:phaseDiagram_spiralPEPS_GSE_TL}].
At this point the system undergoes a transition from a spiral ground state (at finite $\chi_{B}$) with wave vector $\mathbf{k} = (2\pi/3, 2\pi/3)$, found by variational optimization, to a fully translational invariant state with $\mathbf{k} = (0, 0)$. 

As a quantitative check on the variational quality of the optimized states, we have also performed a zero-variance extrapolation using the recently introduced energy-variance calculation for infinite PEPS~\cite{estay2025accurate}, adapted here to the triangular lattice geometry.
This extrapolation is not used to infer the absence of magnetic order: the lowest-energy finite-$\chi_B$ triangular-lattice PEPS wave functions already have vanishing local magnetization throughout the ruby--MLL regime.
The variance analysis provides an independent measure of how close these finite-$\chi_B$ wave functions are to an exact eigenstate.
Applying this extrapolation to the ruby and MLL datasets (see the End Matter section for technical details and also Refs.~\cite{estay2025accurate,Naumann2026} therein) yields
\begin{align}
    \begin{split}
        E_0(\text{Ruby}) &= \num{-0.5634 \pm 0.0001},\\
        E_0(\text{MLL}) &= \num{-0.5308 \pm 0.0007}.
    \end{split}
\end{align}
Interestingly, the fully variational energies at $\chi_B = 4$ are already very close to the final extrapolations, demonstrating the exceptional expressivity of the triangular lattice PEPS ansatz.
Furthermore, an inspection of the nearest-neighbor spin-spin correlations on all MLL bonds reveals an accurate three-fold rotational symmetry in the ground states, as shown in the inset of Fig.~\ref{fig:phaseDiagram_spiralPEPS_GSE_TL}.
One can argue that the precision is striking, given that the TL PEPS does not explicitly enforce point-group symmetries, but rather allows them to emerge naturally by adapting the native underlying lattice geometry.

\begin{figure}[t!]
    \centering
\tikzsetnextfilename{phaseDiagram_spiralPEPS_GSM_TL}
    \begin{tikzpicture}
    
    \begin{axis}[
        xlabel=$J_g$,
        ylabel=$m^2$,
        width=\columnwidth,
        label style={font=\small},
        legend columns=3, 
        legend style={
            font=\footnotesize, 
            at={(0.075, 0.82)},
            anchor=south west
        },
        legend image post style={xscale=0.6},
        tick label style={
            font=\footnotesize,
        },
        yticklabel style={
            /pgf/number format/fixed,
            /pgf/number format/precision=3
        },
        scaled y ticks=false,
        every axis/.append style={thick},
        clip mode=individual,
        /tikz/mark size=1pt,
        xmajorgrids=true,
        ymajorgrids=true,
        xmin=-0.05, xmax=1.55,
        ymin=-0.003, ymax=0.115,
    ]
    \addplot[mathematicaplot4!40!white, mark=square*] table [x=Jg,y=D6] {data/phaseDiagram_spiralPEPS_GSM.txt};
    \addlegendentry{SL $\chi_B = 6$}

    \addplot[mathematicaplot5!40!white, mark=square*] table [x=Jg,y=D7] {data/phaseDiagram_spiralPEPS_GSM.txt};
    \addlegendentry{SL $\chi_B = 7$}

    \addplot[mathematicaplot6!40!white, mark=square*] table [x=Jg,y=D8] {data/phaseDiagram_spiralPEPS_GSM.txt};
    \addlegendentry{SL $\chi_B = 8$}

    \addplot[mathematicaplot1, mark=triangle*] table [x=Jg,y=D2] {data/phaseDiagram_spiralPEPS_GSM_TL.txt};
    \addlegendentry{TL $\chi_B = 2$}
    
    \addplot[mathematicaplot3, mark=triangle*] table [x=Jg,y=D3] {data/phaseDiagram_spiralPEPS_GSM_TL.txt};
    \addlegendentry{TL $\chi_B = 3$}

    \addplot[mathematicaplot2, mark=triangle*] table [x=Jg,y=D4] {data/phaseDiagram_spiralPEPS_GSM_TL.txt};
    \addlegendentry{TL $\chi_B = 4$}
    \end{axis}

\end{tikzpicture}
    \caption{The variational states corresponding to the lowest energies (\text{TL}, $\chi_B=4$) show vanishing magnetization from the ruby lattice to the MLL and beyond. We find a small region of finite canted-$120^\circ$ magnetic order before entering the dimer phase, in which the magnetization vanishes again. We note that both for the SL and TL data, the size of the intermediate ordered phase decreases with increasing bond dimension, raising the question of whether it survives in the infinite-bond-dimension limit.}
\label{fig:phaseDiagram_spiralPEPS_GSM_TL}
\end{figure}
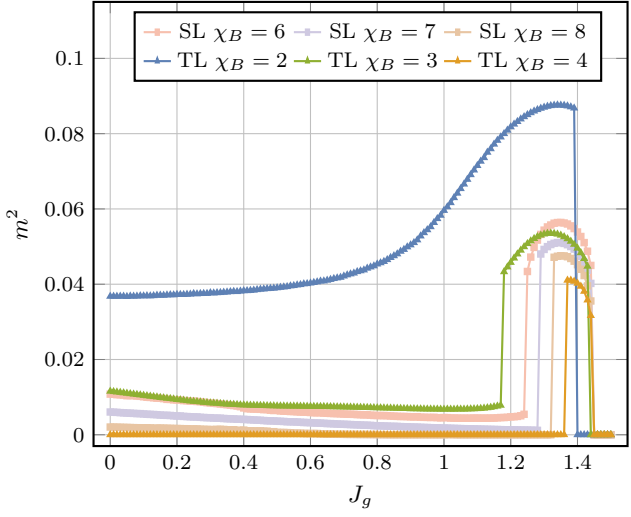

Besides the known transition to the exact dimer phase, both the ground state energy and the spin-spin correlations display an additional kink, indicating the presence of another transition.
To resolve the phase diagram of the model and address the potential presence or absence of magnetic order, we analyze the magnetic order parameter, given by the average staggered magnetization per spin
\begin{align}
    m^2 = \frac{1}{N} \sum_{i=1}^{N} \left( \langle S_i^x \rangle^2 + \langle S_i^y \rangle^2 + \langle S_i^z \rangle^2 \right).
\end{align} 
The results are shown in Fig.~\ref{fig:phaseDiagram_spiralPEPS_GSM_TL}.
For both the infinite PEPS on the SL and TL, we find that increasing the bond dimension suppresses the magnetic order parameter for the entire range from the ruby lattice to the MLL and beyond.
The variational results of lowest energy ($\text{TL},\; \chi_B = 4$) yield no magnetic order for this entire regime.
The magnetization for the infinite PEPS ansatz on the SL, extrapolated against the inverse bond dimension, also yields consistent results of vanishing magnetization.
At smaller bond dimensions, both the SL and TL PEPS ansätze can lock into a canted-$120^\circ$ pattern, whereas increasing the bond dimension suppresses the order parameter.
This behavior suggests that the ordered state is a nearby competing
variational minimum, favored at small bond dimension before the ansatz has sufficient capacity to represent the competing short-range entanglement.
The resulting non-magnetic regime is compatible with earlier indications of strong quantum-disordered tendencies, including one branch of the CCM extrapolations~\cite{Farnell2014}, the pf-FRG indication of an intermediate quantum-disordered regime~\cite{Gresista2023}, and the recent NLCE evidence for a short-range correlated paramagnet at the isolated MLL point~\cite{Schaefer2026}.

At larger values of $J_g$, our variational results consistently reveal a narrow region of finite magnetization separating the disordered and exact dimer phases.
We observe, however, that the extent of this intermediate phase decreases significantly as the bond dimension increases.
Given the fact that, $i)$ our $\chi_B = 4$ TL data is already quite close in energy to the value we obtain from the variance extrapolation at the isotropic points, and $ii)$ the energy in the magnetically ordered phase only decreases moderately with bond dimension, we suspect that the PEPS represent the ground states of the model quite faithfully.
We therefore conjecture this ordered intermediate phase to be present in the ground state phase diagram.
The magnitude of the order parameter might, however, still change as more entanglement can further dress this order with quantum fluctuations.
This leaves open the possibility that the order parameter could ultimately vanish in the true ground state.
We find that the dimer phase consistently emerges for $J_g \gtrsim 1.45$, where the energy approaches $E_0=-3J_g/8$, and the green-bond correlations approach the singlet value $\langle \mathbf S_i \cdot \mathbf S_j \rangle = -3/4$.

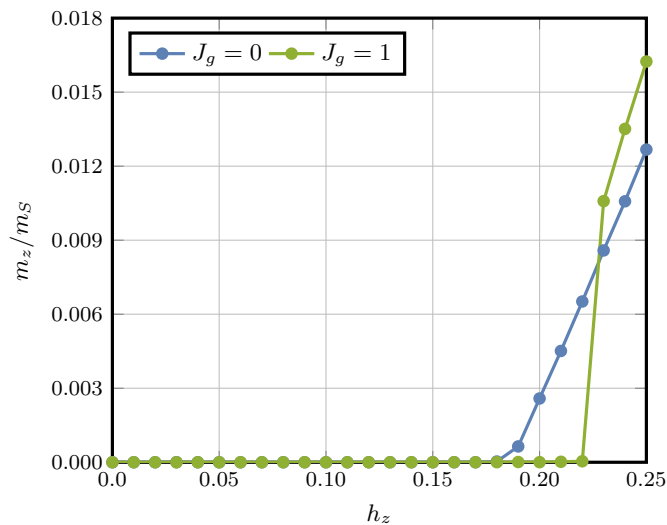
\begin{figure}[ht]
    \centering
    \tikzsetnextfilename{spinGapPlot}
    \begin{tikzpicture}

    \begin{axis}[
        xlabel=$h_z$,
        ylabel=$m_z/m_S$,
        width=\columnwidth,
        legend columns=2,
        legend style={font=\small},
        legend cell align={left},
        label style={font=\small},
        tick label style={font=\footnotesize},
        every axis/.append style={thick},
        /tikz/mark size=1.75pt,
        legend pos=north west,
        every axis/.append style={very thick},
        yticklabel style={
            /pgf/number format/fixed,
            /pgf/number format/precision=3
        },
        scaled y ticks=false,
        xmajorgrids=true,
        ymajorgrids=true,
        xmin=0, xmax=0.25,
        xtick = {0.0, 0.05, 0.10, 0.15, 0.20, 0.25},
        xticklabels = {0.0, 0.05, 0.10, 0.15, 0.20, 0.25},
        ymin=0, ymax=0.018,
        ytick = {0.00, 0.003, 0.006, 0.009, 0.012, 0.015, 0.018},
        yticklabels = {0.000, 0.003, 0.006, 0.009, 0.012, 0.015, 0.018},
    ]
    
    \addplot[mathematicaplot1, mark=*] table [x=Hz,y=Sz] {data/magField_Jg_0.00_TL_chiB_4.txt};
    \addlegendentry{$J_g = 0$}

    \addplot[mathematicaplot3, mark=*] table [x=Hz,y=Sz] {data/magField_Jg_1.00_TL_chiB_4.txt};
    \addlegendentry{$J_g = 1$}
    
    \end{axis}
    
\end{tikzpicture}
    \caption{Average spin component $m_z = \langle S_z \rangle$ normalized to the saturation value $m_S = 1/2$ for the isotropic ruby lattice and MLL, obtained with the lowest-energy TL PEPS at $\chi_B = 4$. Both curves display a zero-magnetization plateau, giving the operational estimates $\Delta_{J_g=0}\simeq 0.17$ and $\Delta_{J_g=1}\simeq 0.22$ for the field scale at which magnetization first appears. The onset is gradual for the ruby lattice and more abrupt for the MLL, indicating different low-lying magnetization processes despite the common absence of zero-field order.}
    \label{fig:magenticfield}
\end{figure}

Given the large size of the newly identified disordered phase between the ruby lattice and the MLL, it is natural to ask whether its properties are continuous throughout or if there are actually multiple different phases.
To pursue this question, we performed simulations with an external magnetic field $0 \le h_z \le 0.25$ for our most accurate variational states ($\text{TL},\; \chi_B = 4$), which yield zero magnetization at zero field, cf. Fig.~\ref{fig:phaseDiagram_spiralPEPS_GSM_TL}.
Although the magnetic field only leaves a remaining $\mathrm{U}(1)$ symmetry in the Hamiltonian, the spiral PEPS ansatz can still be employed by choosing the rotation axis along the direction of the field (see End Matter section for further details).
Focusing on the isotropic ruby lattice and MLL, we find the magnetization curves shown in Fig.~\ref{fig:magenticfield}.

For both isotropic Heisenberg models, the system develops a visible zero-magnetization plateau, resulting in the operational spin-gap estimates $\Delta_{J_g = 0} \sim 0.17$ and $\Delta_{J_g = 1} \sim 0.22$.
Both isotropic endpoints show a finite zero-magnetization plateau within the present TL $\chi_B=4$ simulations.
We therefore interpret the corresponding field scales as estimates of finite triplet gaps in the variational state.
The character of the magnetization onset is nevertheless different at the two endpoints: on the ruby lattice the magnetization grows gradually once the plateau ends, whereas on the MLL it rises more sharply.
This distinction may reflect different low-lying triplet sectors or a change in the dominant short-range correlations, but the present data do not establish a separate phase transition between the two endpoints.
The finite zero-magnetization plateau at the ruby endpoint is especially noteworthy in this context: it shows that the ruby geometry supports a gapped non-magnetic state already in the minimal $SU(2)$-symmetric nearest-neighbor Heisenberg limit, distinct from but complementary to the blockade-derived ruby-lattice models realized in Rydberg arrays.

These observations point to a connected symmetric quantum-paramagnetic regime extending from the ruby lattice to the MLL.
Within the accuracy of our variational calculations, the absence of local magnetization and the recovery of lattice-rotation symmetry exclude the conventional canted-$120^\circ$ ordered state.
They do not, however, determine the nature of the paramagnet.
The remaining possibilities include a gapped $\mathbb Z_2$ QSL, a featureless symmetric short-range paramagnet, or a weak valence-bond state whose order parameter is below the present resolution.
The NLCE results at the MLL point~\cite{Schaefer2026}, which point to resonating hexagonal motifs, are consistent with the weak green-bond correlations found here, but our local observables alone do not select a unique microscopic interpretation.

We have also extensively examined tensor-network diagnostics that are more directly sensitive to possible topological order.
In a PEPS description, $\mathbb Z_2$ topological order is not expected to appear through a conventional local order parameter, but through the virtual symmetry of the tensor and the sector structure of the boundary transfer matrix.
A featureless paramagnet should admit an injective PEPS representation with a single dominant transfer-matrix fixed point, whereas a $\mathbb Z_2$ spin liquid should show stable virtual $\mathbb Z_2$ sectors associated with nontrivial string or flux insertions.
At the bond dimensions reached here, we do not find sector-resolved signatures stable enough to identify the state as intrinsically topologically ordered.
This does not rule out a $\mathbb Z_2$ interpretation: finite bond dimension and boundary-environment truncation may allow local energies and short-range correlations to converge before the virtual-sector structure is fully resolved.
Moreover, recent exact-diagonalization and projected-wave-function results suggest that a gapped $\mathbb Z_2$ state may be more clearly stabilized away from the high-symmetry interpolation line studied here~\cite{Ebert2026}.
We therefore conclude that the present calculations establish an extended gapped non-magnetic regime, while its topological or featureless character remains largely open.\\

\paragraph{Conclusion and outlook.} 
We have presented a variational infinite-PEPS study of the spin-$1/2$ Heisenberg antiferromagnet on the generalized maple-leaf lattice, along the axis connecting the isotropic ruby and maple-leaf limits.
This axis provides a particularly stringent setting for frustrated magnetism: it interpolates between two closely related Archimedean lattices while changing the local connectivity and frustration without introducing disorder, longer-range interactions, or explicit anisotropy.
Using two complementary tensor-network representations---one on the mapped square lattice and one directly on the native triangular lattice---we find that the lowest-energy variational states form an extended non-magnetic regime connecting the ruby and maple-leaf endpoints.

The evidence for this regime is threefold.
First, the local magnetic order parameter is suppressed systematically with increasing bond dimension and vanishes, within numerical accuracy, in the best variational states throughout the interval from the ruby lattice to the maple-leaf lattice and beyond.
Second, the nearest-neighbor bond correlations recover the expected threefold lattice-rotation symmetry without this symmetry being imposed on the tensor.
Thus the absence of magnetic order is not accompanied by an obvious spontaneous lattice-symmetry breaking in the accessible variational manifold.
Third, magnetic-field simulations at the two isotropic endpoints provide an independent probe: both show a zero-magnetization plateau over a finite field interval, with operational scales $\Delta_{J_g=0}\sim0.17$ and $\Delta_{J_g=1}\sim0.22$ in the present TL $\chi_B=4$ calculations.
The ruby endpoint displays a gradual magnetization onset, whereas the maple-leaf endpoint shows a sharper onset, indicating distinct low-lying magnetization processes within a common non-magnetic regime.

Our results place the generalized maple-leaf model among the cleanest nearest-neighbor Heisenberg platforms in which to study how semiclassical order fails in two dimensions.
The key outcome is not merely the absence of a local moment, but the emergence of a broad gapped non-magnetic regime in the thermodynamic limit, bounded by nearby magnetically ordered and dimerized competitors.
The ruby endpoint is particularly notable: the same lattice geometry that underlies recent Rydberg-array routes to topological spin liquids also supports a gapped non-magnetic state in the minimal $SU(2)$-symmetric Heisenberg limit.

We pose the topological character of this regime as a compelling open question.
In PEPS language, a definitive identification of a $\mathbb Z_2$ QSL requires stable virtual $\mathbb Z_2$ sectors, topological transfer-matrix fixed points, or equivalent sector-resolved boundary diagnostics, rather than local observables alone.
The ruby--maple-leaf interpolation therefore provides a concrete target for future topological diagnostics in frustrated Heisenberg models.
More broadly, our results showcase the ability of modern tensor-network methods to access highly frustrated two-dimensional quantum magnets in regimes that remain challenging for alternative numerical approaches.\\

\paragraph{Simulations and data.}
The numerical simulations were performed using the publicly available \emph{variPEPS} library~\cite{naumann_varipeps_python}, as presented in Ref.~\cite{Naumann2023,Naumann2026}.\\

\paragraph{$\text{CO}_2$-emissions table.} 
The TN calculations in this work demanded significant computational resources. To highlight the environmental impact, Table~\ref{tab:carbonEmissions} shows a conservative estimate of the carbon emissions from these simulations, advocating for greater carbon footprint awareness in numerical research.
\begin{table}[h!]
    \centering
    \begin{tblr}{
      colspec = {X r},
      hline{1,Z} = {1pt},
      hline{2} = {0.6pt}
    }
        \textbf{TN simulations} & \\
        Total kernel time & $\sim \qty{11.3}{\million\hour}$ \\
        Thermal design power per kernel & \qty{12}{\watt} \\
        Total energy consumption & $\sim \qty{136}{\mega\watt\hour}$ \\
        Average emission of CO$_2$ in Germany in 2023~\cite{CO2_electricity_germany} & \qty[per-mode = symbol, sticky-per, bracket-unit-denominator = false]{0.38}{\kg\per\kWh} \\
        Total CO$_2$ emission & $\sim \qty{51500}{\kg}$ \\
        Were the emissions offset? & {\textbf{partially}} \\
    \end{tblr}
    \caption{Estimate of the carbon emissions produced by the numerical simulations in this work, calculated according to the Scientific CO$_2$nduct project~\cite{Sweke2022}.}
    \label{tab:carbonEmissions}
\end{table}

\begin{acknowledgments}
\paragraph{Acknowledgments.}
We acknowledge inspiring discussions with L.~Balents, S.~Bhattacharjee, P.~L.~Ebert, A.~Haller, J.~Hasik, F.~Krein, A.~L\"auchli, C.~Liu, A.~Nevidomskyy, A.~Nietner, S.~Parameswaran, K.~Penc, R. Schäfer and A.~Wietek. This work has been funded by the Deutsche Forschungsgemeinschaft (DFG, German Research Foundation) under the project number 277101999 -- CRC 183 (project B01), and the BMBF (MUNIQC-Atoms, FermiQP). P.~S.\ and J.~N.\ thank the ZEDV (IT support) of the physics department, Freie Universität Berlin, for computing time and technical support. In particular, we thank J.~Behrmann, C.~Hoffmann and J.~Dreger. We also acknowledge the computing time provided by the HPC Service of FUB-IT, Freie Universität Berlin~\cite{Bennett2020}. J.~E.\ acknowledges funding of the ERC (DebuQC). The work of Y.~I.\ was performed in part at the Aspen Center for Physics, which is supported by National Science Foundation Grant No.~PHY-2210452. The participation of Y.~I.\ at the Aspen Center for Physics was supported by the Simons Foundation. The research of Y.~I.\ was carried out, in part, at the Kavli Institute for Theoretical Physics in Santa Barbara during the “A New Spin on Quantum Magnets” program in summer 2023 and ``Correlated Gapless Quantum Matter" program in spring 2024, supported by the National Science Foundation under Grant No.~NSF PHY-1748958. Y.~I.\ thanks the Pollica Physics Centre for the workshop ``Exotic quantum matter from quantum spin liquids to novel field theories'' where this manuscript was completed. Y.~I.\ acknowledges support from the ICTP through the Associates Programme, from the Simons Foundation through Grant No.~284558FY19, and IIT Madras through the Institute of Eminence (IoE) program for establishing QuCenDiEM (Project No.~SP22231244CPETWOQCDHOC). Y.~I.\ also acknowledges the use of the computing resources at HPCE, IIT Madras.
\end{acknowledgments}

\bibliography{references}

\newpage

\appendix

\onecolumngrid
\newpage
\section{End Matter}
\twocolumngrid

\subsection{Spiral PEPS setup for the maple-leaf lattice}

The variational PEPS simulations of the maple-leaf lattice are performed using the spiral PEPS ansatz, applied both to infinite PEPS on the square lattice~\cite{Hasik2023} and the triangular lattice~\cite{Naumann2026}.
The full many-body state vector
\begin{align}
    \ket{\psi(A, \mathbf k)} = U(\mathbf k) \sum_{\lbrace s_i \rbrace} C_{\lbrace s_i \rbrace}(A) \ket{\lbrace s_i \rbrace}
\end{align}
is expressed as a network of a single tensor $A$, mapping from parameter space to the physical amplitudes by $C_{\lbrace s_i \rbrace}(A)$, together with a global unitary transformation parameterized by a \emph{wave vector} $\mathbf{k}\in \mathbb{R}^2$.
The global unitary transformation can be decomposed into a product of spatially dependent local unitaries as
\begin{align}
    U(\mathbf k) = \prod_{\mathbf r} u_{\mathbf r}(\mathbf k)
\end{align}
that act only on the combined physical index of the six spins.
For the case of the triangular lattice, the full infinite spiral PEPS state vector is then given by
\begin{align*}
    \begin{split}
        \includegraphics[scale = 1.0]{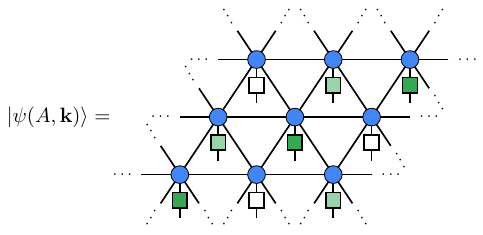}
    \end{split},
\end{align*}
where the spiral unitaries generate a three-sublattice structure and an 18-site unit cell.
One advantage of this approach is that for models with a global symmetry such as $\mathrm{SU}(2)$, arbitrary unit cells can be generated by only a single-site tensor and corresponding relative rotations.
This is computationally more efficient than choosing a large unit cell of different tensors and allows us to reach notably large bond dimensions, given no global symmetries are exploited in the TN~\cite{Silvi2019,Schmoll2020}.
Moreover, the spiral pitch vector of the magnetically ordered state can be variationally optimized together with the PEPS tensor~\cite{Hasik2023}, such that the ansatz faithfully represents the structure of the target state.
This is important because quantum fluctuations lead to a shift of the spiral pitch vector and the relative angle between spins in the unit cell compared to their classical values~\cite{Chubukov-1984,Iqbal-2016}.
The tensor $A$ has four (six) virtual indices with a bulk bond dimension $\chi_B$ on the square (triangular) lattice, as well as a physical index of dimension $d = 2^6$ due to the chosen coarse-graining.
To this end, the basis of six spins is coarse-grained into an effective PEPS site, which introduces next-to-nearest neighbor interactions on the resulting square lattice, while it remains strictly nearest-neighbor on the triangular lattice.
In order to contract the infinite TN for the calculation of its norm and expectation values, we employ a regular \emph{corner transfer matrix renormalization group} (CTMRG) scheme~\cite{Nishino1996,Nishino1997,Orus2009,Naumann2026}.
It calculates fixed-point environment tensors using an iterative power method. 
The unavoidable approximations in this procedure are controlled by an additional refinement parameter, the environment bond dimension $\chi_E$.
In all simulations, $\chi_E$ is chosen to be sufficiently high for the results to be converged, or as the largest value for computational feasibility.

For the labeling convention of the MLL, the two basis vectors are given by
\begin{align}
    \mathbf{a}_1 = -\frac{\sqrt{7}}{2} \begin{pmatrix} 1 \\ \sqrt{3} \end{pmatrix} \hspace{1.5cm} \mathbf{a}_2 = \sqrt{7} \begin{pmatrix} 1 \\ 0 \end{pmatrix}.
\end{align}
Here, the lattice constant is set to $a = 1$.
For the spiral PEPS ansatz on the square lattice, we map these vectors to the orthonormal vectors $\mathbf{a}_1 \mapsto -\hat e_y$ and $\mathbf{a}_2 \mapsto +\hat e_x$.
For the triangular PEPS ansatz instead, the vectors $\mathbf{a}_1$ and $\mathbf{a}_2$ are directly spanning the lattice, but they are not orthogonal to each other.
The spiral is then defined collectively for the six-site basis, still parametrized by a single two-dimensional wave vector $\mathbf{k} \in \mathbb R^2$.
It is incorporated into the calculation of expectation values of vertical, horizontal and diagonal interactions on the square and triangular lattice.
It is advantageous to choose the spiral rotation around the $y$-axis, which allows one to work entirely in the $xz$-plane. 
Here, a PEPS ansatz with real-valued tensor coefficients can be used~\cite{Hasik2023}, which reduces the computation cost compared to a complex-valued simulation.
The local unitary transformations on site $\mathbf{r}$ are then given by
\begin{align}
    u_{\mathbf{r}}(\mathbf k, \mathbf{r}^\prime) = \exp \left\lbrack i\pi (\mathbf k \cdot \mathbf{r}^\prime) S_{\mathbf{r}}^{y} \right\rbrack.
\end{align}
Due to the $\mathrm{SU}(2)$ symmetry of the Heisenberg Hamiltonian, its action on the interaction terms only depend on the relative position of the coarse-grained sites involved, so that we have $\mathbf{r}^\prime = \mathbf{a}_1$, $\mathbf{r}^\prime = \mathbf{a}_2$ and $\mathbf{r}^\prime = \mathbf{a}_1 + \mathbf{a}_2$ for the three types of interactions (vertical, horizontal and diagonal).
The eighteen site unit cell of the classical ground state then corresponds to a wave vector $\mathbf{k} = (2\pi/3, 2\pi/3)$ on the coarse-grained lattice. 
This is also the structure we found by numerically optimizing over $\mathbf{k}$ for the ground state outside the exact dimer phase.
Note that if the structure of the ground state is known, a fixed wave vector $\mathbf{k}$ can also be imprinted, leaving only the bulk PEPS tensor to be variationally optimized.

When including an external magnetic field, the Hamiltonian only possesses a $\mathrm{U}(1)$ symmetry.
Fortunately, the spiral PEPS ansatz can still be used with a rotation around the axis of the field.
However, in contrast to the $\mathrm{SU}(2)$-symmetric Hamiltonian, now a PEPS tensor with complex coefficients needs to be chosen.
The wave vector can be optimized to find the correct structure for the ground states, which allows the spiral PEPS ansatz to capture the correct pattern for magnetization plateaus, as for instance found for the isotropic Heisenberg model on the maple-leaf lattice at $\langle S_z \rangle = 0$.
\begin{figure}[ht]
    \centering
    \includegraphics[width = 0.45\columnwidth]{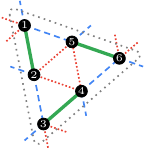}
    \caption{Definition of one unit cell, the six-site basis of the maple-leaf lattice.}
    \label{fig6}
\end{figure}
The basis of the maple-leaf lattice is defined in Fig.~\ref{fig6}.
In the translational invariant spiral PEPS ansatz there are only six individual spins, and fifteen different nearest-neighbor bonds in the network.
The unit cell sites are spanned by the following six basis vectors
\begin{align}
    \begin{aligned}
    \mathbf{b}_1 &= \begin{pmatrix} 0 \\ 0 \end{pmatrix}, & \mathbf{b}_4 &= \frac{1}{\sqrt{7}} \begin{pmatrix} 3 \\ -2\sqrt{3} \end{pmatrix}, \\
    \mathbf{b}_2 &= \frac{1}{2\sqrt{7}} \begin{pmatrix} 1 \\ -3\sqrt{3} \end{pmatrix}, & \mathbf{b}_5 &= \frac{1}{2\sqrt{7}} \begin{pmatrix} 5 \\ -\sqrt{3} \end{pmatrix} ,\\
    \mathbf{b}_3 &= \frac{1}{\sqrt{7}} \begin{pmatrix} 1 \\ -3\sqrt{3} \end{pmatrix} ,& \mathbf{b}_6 &= \frac{1}{\sqrt{7}} \begin{pmatrix} 5 \\ -\sqrt{3} \end{pmatrix}.
    \end{aligned}
\end{align}
Those basis vectors do not enter in the unitary transformation of the spiral PEPS ansatz, as it only depends on the two-dimensional wave vector $\mathbf{k}$.
However, they are important in the calculation of structure factors, as described below.

\subsection{Energy variance extrapolations}

\begin{figure}[ht]
    \centering
    \begin{minipage}{0.95\columnwidth}
        \centering
        \begin{tikzpicture}

		\begin{axis}[
			xlabel=$\sigma^2$,
			ylabel=$E_0$,
			width=0.95\columnwidth,
			legend style={
				at={(0.025, 0.975)},
				anchor=north west
			},
			legend image post style={xscale=0.6},
			legend columns=1,
			legend cell align={left},
			label style={font=\small},
			every axis/.append style={thick},
			/tikz/mark size=1.25pt,
			xmajorgrids=true,
			ymajorgrids=true,
			xmin=0.00, xmax=0.25,
			x tick label style={
				/pgf/number format/.cd, 
					fixed,
					fixed zerofill,
					precision=2,
				/tikz/.cd 
			},
			y tick label style={
				/pgf/number format/.cd, 
					fixed,
					fixed zerofill,
					precision=3,
				/tikz/.cd 
			},
		]

		\addplot[
			color = mathematicaplot1, 
			mark = square, 
			line width=1.5pt, 
			mark size=1.5pt,
			only marks,
		]
		coordinates {
			(0.22751720, -0.5546943245944201) 
			(0.10004001, -0.5599207913184876) 
			(0.00809683, -0.5631201069578108)
		};
		\addlegendentry{$J_g = 0.00$};

        \addplot [
			domain=0:0.25,
			samples=100,
			color=black,
			thick,
		]
		{-0.5633789523460943 + 0.03173985099071113 * x^1 + 0.028267977148656236 * x^2};
		
		\end{axis}

	\end{tikzpicture}
    \end{minipage}
    \begin{minipage}{0.95\columnwidth}
        \centering \begin{tikzpicture}

    \begin{axis}[
        xlabel=$\sigma^2$,
        ylabel=$E_0$,
        width=0.95\columnwidth,
        legend style={
            at={(0.025, 0.975)},
            anchor=north west
        },
        legend image post style={xscale=0.6},
        legend columns=1,
        legend cell align={left},
        label style={font=\small},
        every axis/.append style={thick},
        /tikz/mark size=1.25pt,
        xmajorgrids=true,
        ymajorgrids=true,
        xmin=0.00, xmax=0.16,
        x tick label style={
            /pgf/number format/.cd, 
                fixed,
                fixed zerofill,
                precision=2,
            /tikz/.cd 
        },
        y tick label style={
            /pgf/number format/.cd, 
                fixed,
                fixed zerofill,
                precision=3,
            /tikz/.cd 
        },
    ]

    \addplot[
        color = mathematicaplot1, 
        mark = square, 
        line width=1.5pt, 
        mark size=1.5pt,
        only marks,
    ]
    coordinates {
        (0.14401596, -0.5176101230885294) 
        (0.07921056, -0.5263941358230954) 
        (0.01638432, -0.5304384142296198) 
    };
    \addlegendentry{$J_g = 1.00$};

    
    \addplot [
        domain=0:0.16,
        samples=100,
        color=black,
        thick,
    ]
    {-0.5307694064798351 + 0.011065280055048378 * x^1 + 0.5576361475713136 * x^2};


    
    \end{axis}

\end{tikzpicture}
    \end{minipage}
    \caption{Ground state energy $E_0$ versus energy variance $\sigma^2$, and extrapolation to the zero-variance limit for the isotropic Heisenberg model on the ruby ($J_g = 0.00$) and maple-leaf lattice ($J_g = 1.00$). Due to the different amounts of geometric frustration, the extrapolation is more linear for the ruby and curved for the maple-leaf point. To estimate an error to the extrapolated value, we use a linear fit of the two largest data points.}
    \label{fig:variance_extrapolation}
\end{figure}
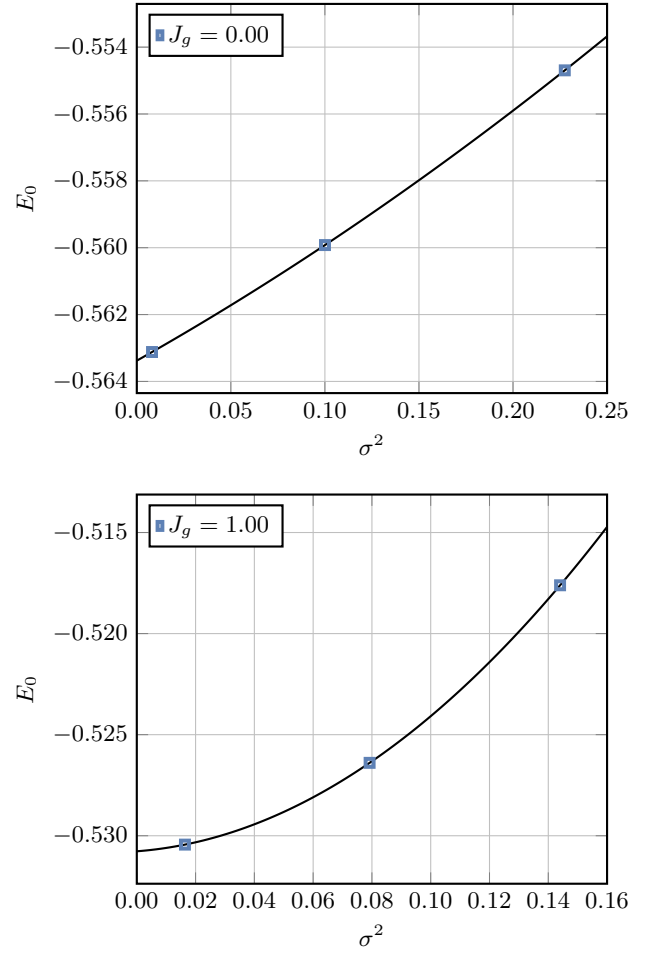

Based on the proposal in Ref.~\cite{estay2025accurate} and significant structural modifications for triangular lattice PEPS~\cite{Naumann2026}, the energy variance is computed using a large-cell CTMRG algorithm for a finite patch of radius $L$ along the $\mathbf{a}_1$, $\mathbf{a}_2$ and $\mathbf{a}_1 + \mathbf{a}_2$ lattice directions.
Using a maximal value of $L = 10$ (corresponding to a maximal elongation of $2L+2$ sites along to the centered bond), the energy variance is reasonably well converged in the system size, even for the most accurate $\chi_B = 4$ variational results.
The extrapolation is performed for the isotropic ruby and maple-leaf lattice limits, i.e., for $J_g = 0$ and $J_g = 1$, and is shown in Fig.~\ref{fig:variance_extrapolation}.
The significantly larger amount of geometric frustration on the maple-leaf lattice leads to a stronger curvature of the data points, as compared to the almost linear behavior on the ruby lattice. 
Still, the triangular lattice PEPS ansatz represents an excellent approximation to the ground states of both models, indicating by the close proximity of the lowest variational energies at $\chi_B = 4$ and the extrapolated values.
Realistically, a larger PEPS bond dimension would only lead to a marginal improvement of the ground state approximation, certainly at least for the isotropic ruby lattice.
Using a linear fit of the two largest data points, we can estimate an error to the zero-variance extrapolated value.

\newpage

\onecolumngrid
\section{Supplemental Material}
\twocolumngrid

\subsection{Virtual $\mathbb{Z}_2$ symmetry in the PEPS simulations}

Tensor networks provide a significant advantage in numerical simulations by allowing the incorporation of global symmetries. 
Each tensor index is assigned quantum numbers corresponding to the underlying symmetry group, with internal couplings dictated by the appropriate fusion rules. 
This results in a sparse block structure for each tensor, effectively reducing the number of variational parameters in the ansatz. 
While this generally leads to improvements in the computational cost, it also allows targeting specific symmetry sectors for the overall wave function, and thereby a symmetry-resolved analysis. 

Motivated by the possibility of a quantum spin liquid ground state for the Heisenberg model on the isotropic maple-leaf lattice, here we want to exploit the $\mathbb Z_2$ symmetry in the variational PEPS simulations.
Each tensor index, typically associated with a complex vector space $\mathbb V = \mathbb C^{a}$, where $a = [p^6, \chi_B, \chi_E]$ in our simulations, is now represented as 
\begin{align}
    \mathbb V = \bigoplus_{q = \{0, 1\}} d_q \mathbb V_q,
\end{align}
where the quantum numbers can only be the trivial and the non-trivial irreps, $q = 0$ and $q = 1$, respectively.
The physical index of the coarse-grained PEPS tensor is then given by
\begin{align}
    \mathbb V_p = \left(0_1 \oplus 1_1 \right)^{\otimes 6} = (0_{32} \oplus 1_{32}).
\end{align}
However, then implementing the $\mathbb Z_2$ symmetry in the usual way, we are facing two problems: $(i)$ the spiral PEPS ansatz can no longer be used because the local unitary transformations are non-symmetric, and $(ii)$ for the potential $\mathbb Z_2$ QSL ground state, the symmetry is expected to emerge only at the virtual indices of the PEPS tensor without acting globally as described above.
The first point can be mitigated by enlarging the unit cell back to 18 sites, albeit at a significantly higher computational cost, that is prohibitively large to reach $\chi_B = 8$ as for the spiral ansatz.
However, we can eliminate both obstacles by a simple change in the network.
Instead of representing the physical index in a symmetry-preserving form, we represent the physical Hilbert space according to
\begin{align*}
    \begin{split}
        \includegraphics[scale = 1.0]{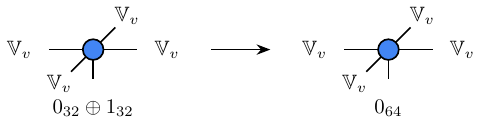}
    \end{split}.
\end{align*}
Since we only use the trivial irrep $q = 0$, the unitary spiral transformations can again be done in its degenerate space, and the $\mathbb Z_2$ symmetry acts entirely on the virtual indices.

\subsection{Calculation of structure factors}

\begin{figure}[t]
    \centering
    \includegraphics[scale = 1.0]{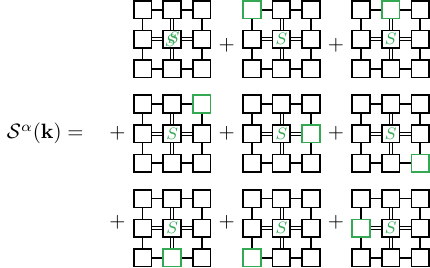}
    \caption{Calculation of one component of the static spin structure factor in Eq.~\eqref{eq:staticSpinStructureFactor_1}. Each green tensor contains all phases and the action of the spin operators arising from the respective semi-infinite part of the square lattice. In the first diagram, both spin operators act on the same site, while in the remaining ones there is only one local operator.}
    \label{fig7}
\end{figure}

In this section we will outline the calculation of PEPS structure factors, based on a modified CTMRG summation scheme~\cite{corboz16_variat_optim_with_infin_projec,ponsioen20_excit_with_projec_entan_pair}.
While this scheme is approximate and more refined schemes have been proposed~\cite{Ponsioen2023, PRXQuantum.5.010335}, the short correlation lengths and large coarse-grained six-site basis justify its use.
We first illustrate the procedure for a regular square lattice, and after that highlight the modifications for non-Bravais lattices such as the MLL.
Exploiting translational invariance of the PEPS ansatz, the square lattice structure factor is given by
\begin{align}
    \begin{split}
        \mathcal{S}({\mathbf{k}}) &= \sum_{\alpha} \mathcal{S}^\alpha(\mathbf{k}) \\ 
        &= \sum_{\alpha} \sum_{x,y} e^{\mathrm{i} (k_x x + k_y y)} \langle \Psi \vert S_{(x,y)}^\alpha \cdot S_{(0,0)}^\alpha \vert \Psi \rangle,
    \end{split}
    \label{eq:staticSpinStructureFactor_1}
\end{align}
where $\alpha$ runs over all spin components $\lbrack x, y, z \rbrack$.
Each part $\mathcal{S}^\alpha(\mathbf{k})$ can be computed by summing up different tensor network diagrams, as shown in Fig.~\ref{fig7}.
In each diagram, the phases $\exp({\mathrm{i}(k_x x + k_y y)})$ and the action of the spin operators $S^{\alpha}$ at position $(x,y)$ are contained in modified CTMRG environment tensors, shown in green. 
They collect the individual contributions of the structure factor arising from the respective semi-infinite part of the square lattice.
The remaining black environment tensors are those for the norm of the infinite PEPS.
The local tensor contains either the effect of both spin operators at the same position $(0,0$), or a single spin operator at this position.
The phases and spin operators for each lattice site $(x,y)$ can be accounted for in a modified CTMRG absorption routine.
To this end, each CTMRG run computes two sets of environment tensors, one for the norm of the quantum state and one that contains the structure factor components.
For the directional CTMRG we make the (somewhat arbitrary) choice, that a bottom absorption is along the lattice vector $\mathbf{a}_1$ and a right absorption is along the lattice vector $\mathbf{a_2}$.
Consequently, a top absorption is performed along $-\mathbf{a}_1$ and a left absorption along $-\mathbf{a}_2$.
Focusing on a left absorption move, the regular edge tensor $T$ is updated by 
\begin{align*}
    \begin{split}
        \includegraphics[scale = 1.0]{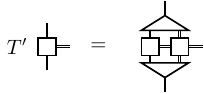}
    \end{split}.
\end{align*}
While absorbing a PEPS tensor and its conjugate, projectors are required to reduce the bond dimension back to $\chi_E$ by restricting to the most important subspace.
The corner tensors $C$ are updated in a similar fashion, absorbing an edge tensor and using truncation projectors, i.e.,
\begin{align*}
    \begin{split}
        \includegraphics[scale = 1.0]{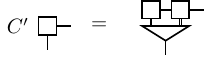}
    \end{split}.
\end{align*}
For the edge tensor $T_p$ containing the phases and spin operators, the absorption is given by
\begin{align*}
    \begin{split}
        \includegraphics[scale = 1.0]{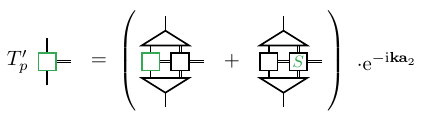}
    \end{split}.
\end{align*}
In the first part a regular PEPS tensor is absorbed, thereby shifting all operators already contained in $T_p$, one lattice site to the left.
In the second part, a new spin operator $S^{\alpha}$ is absorbed into the left boundary tensor $T$.
When weighted with the corresponding phase, the tensor $T_p$ will converge to a collection of terms
\begin{align}
    T_p \sim \hdots + \mathrm{e}^{-\mathrm{i}2k_y} S^{\alpha}_{(-2, 0)} + \mathrm{e}^{-\mathrm{i}k_y} S^{\alpha}_{(-1, 0)},
\end{align}
which generates all structure factor contributions for the semi-infinite row left of site $(0,0)$.
The corner tensors containing phases and operators is updated by
\begin{align*}
    \includegraphics[scale = 1.0]{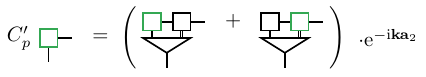}.
\end{align*}
Since every corner participates in two perpendicular absorption steps, it collects all structure factor contributions in one semi-infinite corner of the network.
Let us note, that there are several subtleties in the CTMRG routine, especially when used for a non-trivial PEPS unit cell~\cite{Naumann2023}.
For the environment tensors containing phase factors and spin operators, it is also crucial to normalize them with respect to the environment tensors for the norm of the quantum state, so that the diagrams in Fig.~\ref{fig7} are not skewed.

The extension to non-Bravais lattices such as the MLL is now rather straightforward.
Due to the non-trivial basis, the structure factor in Eq.~\eqref{eq:staticSpinStructureFactor_1} needs two additional sums over the basis sites and relative phase factors between them.
It is given by
\begin{align}
    \begin{aligned}
        \mathcal{S}({\mathbf{k}}) &= \sum_{i,j} \sum_{m,n} e^{\mathrm{i} \mathbf{k} \cdot (\mathbf{R}_i - \mathbf{R}_j)}  e^{\mathrm{i} \mathbf{k} \cdot (\mathbf{b}_m - \mathbf{b}_n)} \\
        & \times \left\langle \mathbf{S}(\mathbf{R}_i + \mathbf{b}_m) \cdot \mathbf{S}(\mathbf{R}_j + \mathbf{b}_n) \right\rangle.
    \end{aligned}
    \label{eq:staticSpinStructureFactor_2}
\end{align}
For the specific case of the MLL with a basis of $N_B = 6$ sites, the first TN diagram in Fig.~\ref{fig7} now contains local correlations within the cluster
\begin{align*}
    \begin{split}
        \includegraphics[scale = 1.0]{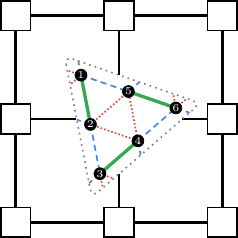}
    \end{split},
\end{align*}
with a total of $N_B \cdot N_B$ different spin-spin correlations, including phases $\exp(\mathrm{i} \mathbf{k} \cdot (\mathbf{b}_m - \mathbf{b}_n))$. 
The remaining diagrams are also affected by one of the loops over the basis sites with phases $\exp(-\mathrm{i} \mathbf{k} \mathbf{b}_n)$.
The second one however appears in the update of the $T_p$ tensors, which now have to include \emph{relative} phases and the action of the spin operator on all basis sites.
For a generic non-Bravais lattice, that can be treated with a CTMRG routine on a square lattice, the generalized absorption is given by
\begin{align}
    \begin{split}
        \includegraphics[scale = 1.0]{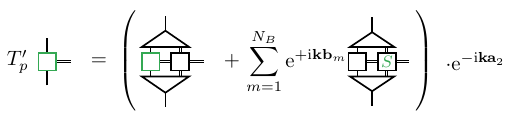}
    \end{split}.
\end{align}
The spin operator $S^\alpha$ needs to be applied to the $m$-th basis site of the local PEPS tensor in the summation.
The update of the corner tensors $C_p$ remains unchanged.

\subsection{Structure factor analysis}

\begin{figure}[ht]
    \centering
    \tikzsetnextfilename{structure_factor}
    \begin{tikzpicture}
    \begin{axis}[
        view={0}{90},
        xlabel=$k_x/\pi$,
        ylabel=$k_y/\pi$,
        width=0.58\columnwidth,
        height=0.58\columnwidth,
        legend style={font=\scriptsize},
        label style={font=\small},
        tick label style={font=\footnotesize},
        every axis/.append style={thick}, 
        colorbar horizontal,
        colorbar style={
            yticklabel style={
                /pgf/number format/fixed,
                /pgf/number format/precision=3
            },
            scaled y ticks=false,
            width=0.87\columnwidth,
            height=1em, 
            xtick={0,1},
            xticklabels={$0$, $\textrm{max}$},
       }, 
       colormap name=bluewhite,
       point meta min=0, point meta max=1,
       xlabel shift=-5,
       ylabel shift=-6,
    ]

    \addplot3[surf,shader=interp,faceted color=none] table[x=kx,y=ky,z=sum] {data/structure_factor_hamVars_+1.00_+0.00_+1.00_+0.000.txt};

    \addplot3[contour lua={labels=false, corners=true, draw color=darkgray!75!gray, number=8}] table[x=kx,y=ky,z=sum] {data/structure_factor_hamVars_+1.00_+0.00_+1.00_+0.000.txt};



    \addplot3[mathematicaplot2, thick, -] coordinates {
    (0.251976315339485, -0.436435780471985, 1)
    (0.50395263067897, 0, 1)
    (0.251976315339485, 0.436435780471985, 1)
    (-0.251976315339485, 0.436435780471985, 1)
    (-0.50395263067897, 0, 1)
    (-0.251976315339485, -0.436435780471985, 1)
    (0.251976315339485, -0.436435780471985, 1)
    };

    \addplot3[mathematicaplot11, thick, -] coordinates {
    (0.251976315339485, -1.30930734141595, 1)
    (1.25988157669742, -0.436435780471985, 1)
    (1.00790526135794, 0.87287156094397, 1)
    (-0.251976315339485, 1.30930734141595, 1)
    (-1.25988157669742, 0.436435780471985, 1)
    (-1.00790526135794, -0.87287156094397, 1)
    (0.251976315339485, -1.30930734141595, 1)
    };

    
    \end{axis}

    \begin{axis}[
        view={0}{90},
        xlabel=$k_x/\pi$,
        width=0.58\columnwidth,
        height=0.58\columnwidth,
        legend style={font=\scriptsize},
        label style={font=\small},
        tick label style={font=\footnotesize},
        every axis/.append style={thick},
       colormap name=bluewhite,
       point meta min=0, point meta max=1.0,
       xshift=0.47\columnwidth,
       xlabel shift=-5,
    ]

    \addplot3[surf,shader=interp,faceted color=none] table[x=kx,y=ky,z=sum] {data/structure_factor_hamVars_+1.00_+1.00_+1.00_+0.000.txt};

    \addplot3[contour lua={labels=false, corners=true, draw color=darkgray!75!gray, number=8}] table[x=kx,y=ky,z=sum] {data/structure_factor_hamVars_+1.00_+1.00_+1.00_+0.000.txt};

    \addplot3[mathematicaplot2, thick, -] coordinates {
    (0.251976315339485, -0.436435780471985, 1)
    (0.50395263067897, 0, 1)
    (0.251976315339485, 0.436435780471985, 1)
    (-0.251976315339485, 0.436435780471985, 1)
    (-0.50395263067897, 0, 1)
    (-0.251976315339485, -0.436435780471985, 1)
    (0.251976315339485, -0.436435780471985, 1)
    };

    \addplot3[mathematicaplot11, thick, -] coordinates {
    (0.251976315339485, -1.30930734141595, 1)
    (1.25988157669742, -0.436435780471985, 1)
    (1.00790526135794, 0.87287156094397, 1)
    (-0.251976315339485, 1.30930734141595, 1)
    (-1.25988157669742, 0.436435780471985, 1)
    (-1.00790526135794, -0.87287156094397, 1)
    (0.251976315339485, -1.30930734141595, 1)
    };


    
    \end{axis}
    \end{tikzpicture}
    \caption{Equal-time structure factors $\mathcal{S}(\mathbf k)$ for the ruby lattice at \mbox{$J_g = 0$} (left) and MLL at \mbox{$J_g = 1$} (right). The first and extended Brillouin zones are indicated in orange and red, respectively.}
    \label{fig5}
\end{figure}
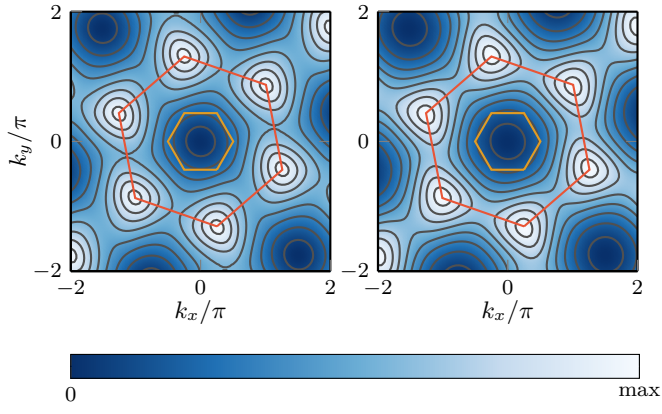

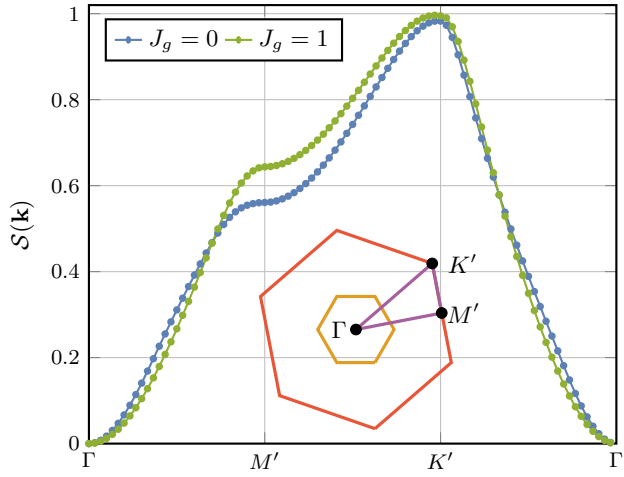
\begin{figure}[htb]
    \centering
    \tikzsetnextfilename{structure_factor_cut}
    \begin{tikzpicture}
    \begin{axis}[
        ylabel=$\mathcal{S}(\mathbf{k})$,
        width=.99\columnwidth,
        legend style={font=\small},
        legend image post style={xscale=0.6},
        legend columns=2,
        legend cell align={left},
        label style={font=\small},
        tick label style={font=\footnotesize},
        every axis/.append style={thick},
        /tikz/mark size=1pt,
        legend pos=north west,
        xmajorgrids=true,
        ymajorgrids=true,
        scaled x ticks=false,
        xticklabel style={
            /pgf/number format/fixed,
            /pgf/number format/precision=3
        },
        scaled y ticks=false,
        xtick={0, 30, 60, 90},
        xticklabels={$\Gamma$, $M^\prime$, $K^\prime$, $\Gamma$},
        xmin = 0, xmax = 90,
        ymin = 0, ymax = 1.02
    ]
    \addplot[mathematicaplot1, mark=*] table [x=N, y=sum] {data/structure_factor_hamVars_+1.00_+0.00_+1.00_+0.000_cut.txt};
    \addlegendentry{$J_g = 0$}

    \addplot[mathematicaplot3, mark=*] table [x=N, y=sum] {data/structure_factor_hamVars_+1.00_+1.00_+1.00_+0.000_cut.txt};
    \addlegendentry{$J_g = 1$}
    \end{axis}

    \begin{axis}[
        view={0}{90},
        width=0.6\columnwidth,
        height=0.6\columnwidth,
       point meta min=0, point meta max=1.25,
       xshift=0.2\columnwidth,
       yshift=-0.034\columnwidth,
       hide axis,
       xmin=-1.8, xmax=1.8,
       ymin=-1.8, ymax=1.8
    ]

    \addplot3[mathematicaplot2, very thick, -] coordinates {
    (0.251976315339485, -0.436435780471985, 1)
    (0.50395263067897, 0, 1)
    (0.251976315339485, 0.436435780471985, 1)
    (-0.251976315339485, 0.436435780471985, 1)
    (-0.50395263067897, 0, 1)
    (-0.251976315339485, -0.436435780471985, 1)
    (0.251976315339485, -0.436435780471985, 1)
    };

    \addplot3[mathematicaplot11, very thick, -] coordinates {
    (0.251976315339485, -1.30930734141595, 1)
    (1.25988157669742, -0.436435780471985, 1)
    (1.00790526135794, 0.87287156094397, 1)
    (-0.251976315339485, 1.30930734141595, 1)
    (-1.25988157669742, 0.436435780471985, 1)
    (-1.00790526135794, -0.87287156094397, 1)
    (0.251976315339485, -1.30930734141595, 1)
    };

    \addplot3[mathematicaplot9, very thick, -] coordinates {
    (0, 0, 1)
    (1.13389341902768, 0.218217890235992, 1)
    (1.0079052613579, 0.872871560943970, 1)
    (0, 0, 1)
    };

    \addplot3[mark=*] coordinates {(0,0,1)} node[label={[label distance=-3]180:$\Gamma$}]{} ;
    \addplot3[mark=*] coordinates {(1.13389341902768, 0.218217890235992, 1)} node[label={[label distance=-4.5]0:$M^\prime$}]{} ;
    \addplot3[mark=*] coordinates {(1.0079052613579, 0.872871560943970, 1)} node[label={[label distance=-0.5]0:$K^\prime$}]{} ;
    \end{axis}
    \end{tikzpicture}
    \caption{Structure factor calculated along the path inside the extended Brillouin zone from the $\Gamma$ point to $M^\prime$, $K^\prime$ and back to the $\Gamma$ point (purple path in the inset). In orange the Brillouin zone and in red the extended Brillouin zone are indicated.}
    \label{fig8}
\end{figure}

We present the equal-time spin structure factors for the isotropic ruby lattice and MLL in Fig.~\ref{fig5} which show broad similarities, being peaked at the $K$-points of the extended Brillouin zone. 
To highlight the differences in the structure factors, we present a plot of the path in the extended Brillouin zone in Fig.~\ref{fig8}, traversing from $\Gamma \mapsto M^\prime \mapsto K^\prime \mapsto \Gamma$. Both curves follow a similar profile, with only moderate differences in slope and absolute value. This is intriguing given the fact that the ground states of the isotropic ruby and maple-leaf lattices exhibit profoundly different characteristics (i.e., gapless vs. gapped).
%

\end{document}